\documentclass[5p,twocolumn]{elsarticle}

\usepackage{graphicx}
\usepackage{dcolumn}
\usepackage{pifont}
\usepackage{bm}
\usepackage{multirow}
\usepackage{amsmath}
\usepackage{float}
\usepackage{txfonts}
\usepackage[version=3]{mhchem}

\usepackage[usenames,dvipsnames]{xcolor}
\definecolor{blue}{RGB}{0,0,225}
\definecolor{cream}{RGB}{222,217,201}
\definecolor{red}{RGB}{225,0,0}

\journal{Elsevier}

\begin{document}
\title{Interface Engineering in Hybrid Iodide \ce{CH3NH3PbI3} Perovskite Using Lewis Base and Graphene towards High Performance Solar Cells}

\author[kimuniv-m]{Chol-Jun Yu\corref{cor}}
\cortext[cor]{Corresponding author}
\ead{cj.yu@ryongnamsan.edu.kp}
\author[kimuniv-m]{Yun-Hyok Kye}
\author[kimuniv-m]{Un-Gi Jong}
\author[kimuniv-l]{Song-Guk Ko}
\author[kimuniv-m]{Kum-Chol Ri}
\author[kimuniv-m]{Song-Hyok Choe}
\author[kimuniv-m]{Jin-Song Kim}
\author[kimuniv-l]{Gwon-Il Ryu}
\author[kimuniv-l]{Byol Kim}

\address[kimuniv-m]{Chair of Computational Materials Design, Faculty of Materials Science, Kim Il Sung University, Ryongnam-Dong, Taesong District, Pyongyang, Democratic People's Republic of Korea}
\address[kimuniv-l]{Chair of Biophysics, Faculty of Life Science, Kim Il Sung University, Ryongnam-Dong, Taesong District, Pyongyang, Democratic People's Republic of Korea}

\begin{abstract}
Perovskite solar cells have achieved a substantial breakthrough via advanced interface engineerings.
Reports have emphasized that combining the hybrid perovskites with Lewis base and graphene improve the performance; the underlying mechanisms are not yet fully understood.
Here, using density functional theory, we show that upon the formation of \ce{CH3NH3PbI3} interfaces with three different Lewis base molecules and graphene, the binding strength with $S$-donors thiocarbamide and thioacetamide is higher than with $O$-donor dimethyl sulfoxide, while the interface dipole and work function reduction tend to increase from $S$-donors to $O$-donor.
Furthermore, we provide evidences of deep trap states elimination in the $S$-donor perovskite interfaces through the analysis of defect formation on \ce{CH3NH3PbI3}(110) surface, and of stability enhancement by estimating activation barriers for iodine atom migrations.
These theoretical predictions are in line with the experimental observation of performance enhancement in the perovskites prepared using thiocarbamide.
\end{abstract}

\maketitle

Perovskite solar cells (PSC) using the emerging class of halide perovskites have provoked a worldwide upsurge of research interest during the past decade, due to a rapid rise in their power-conversion efficiencies (PCE) beyond 24\% by now at half of the price compared to silicon solar cells~\cite{Ono17ael, Li18nrm, YLi18jmca}.
In the structural characteristics of PSCs, there exist several interfaces such as between the perovskite active layer (PAL) and the electron/hole transport layers (ETL/HTL).
Therefore, significant research efforts have been devoted to the interface engineering to achieve a substantial breakthrough in photovoltaics, {\it via} surface passivation~\cite{Schulz18ael, Naghadeh18jpcc, SYe17jacs, Shen18cc, Stolterfoht18ne, Wen16jmca}, compositional tuning~\cite{Qian18apl, HLi18jmca, Lee16acr, Christians18ne} and band alignment alteration~\cite{Ono16jpcl, SZhang18aami, Qiu18jpcb, Lindblad14jpcl, Roiati14nl}.
For the case of organic-inorganic hybrid halide perovskites (e.g., \ce{CH3NH3PbI3} or MAPI), there exist under-coordinated ions such as \ce{Pb^2+} and \ce{I-} on the surface of PALs, which interact with the outdoor \ce{H2O} or \ce{O2} molecules, leading to a hydrolysis of perovskites.
Moreover, these ions, with other surface defects, can act as non-radiative recombination centers for electrons and holes, {\it i.e.} surface trap states, and the trapped charges at interfaces can induce the degradation of PSCs~\cite{Uratani17jpcl}.
Therefore, surface passivation strategy is needed to reduce the trap states of PALs~\cite{XZheng17ne,Grancini17nc}.
To date, several kinds of efficient trap state passivators have been developed, including fullerene~\cite{QWang14ees,YShao14nc,JXu15nc}, graphene derivatives~\cite{Hadadian16am,JTWang16ees}, iododentafluorobenzene~\cite{Abate14nl,Noel14an}, and hole-conductor CuSCN~\cite{SYe16am}.
Furthermore, Lewis bases, such as thiocarbamide (or thiourea) and its derivatives Cu(thiourea)I~\cite{Ko19semsc,Hsieh18ra,SYe17jacs}, were found to be effective for passivating the trap states on the surface of PAL via Lewis acid-base adduct formation with the under-coordinated halide anions and metal cations, where Pb halides (\ce{PbX2}) are known to be Lewis acids.
Considering that the Pb halides are able to form 1:1, 1:2, or 2:1 adducts with monodentate and bidentate ligands with oxygen, sulfur, and nitrogen donors ($O$-, $S$-, and $N$-donor), Lee et al.~\cite{Lee16acr} performed a comparative study of Lewis bases, $O$-donor $N$,$N$-dimethyl sulfoxide (DMSO) and $S$-donor thiocarbamide (TCA), interacting with MAPI and formamidinium lead iodide (\ce{HC(NH2)2PbI3} or FAPI).
It was also suggested that these Lewis bases establish the bulk heterojunction with perovskite, improving the interfacial contacts between the grains~\cite{Ko19semsc}.
In spite of such successes, the origin of performance enhancement by Lewis acid-base interaction has not yet been fully understood.

\begin{figure*}[!th]
\centering
\begin{tabular}{cc}
\includegraphics[clip=true,scale=0.11]{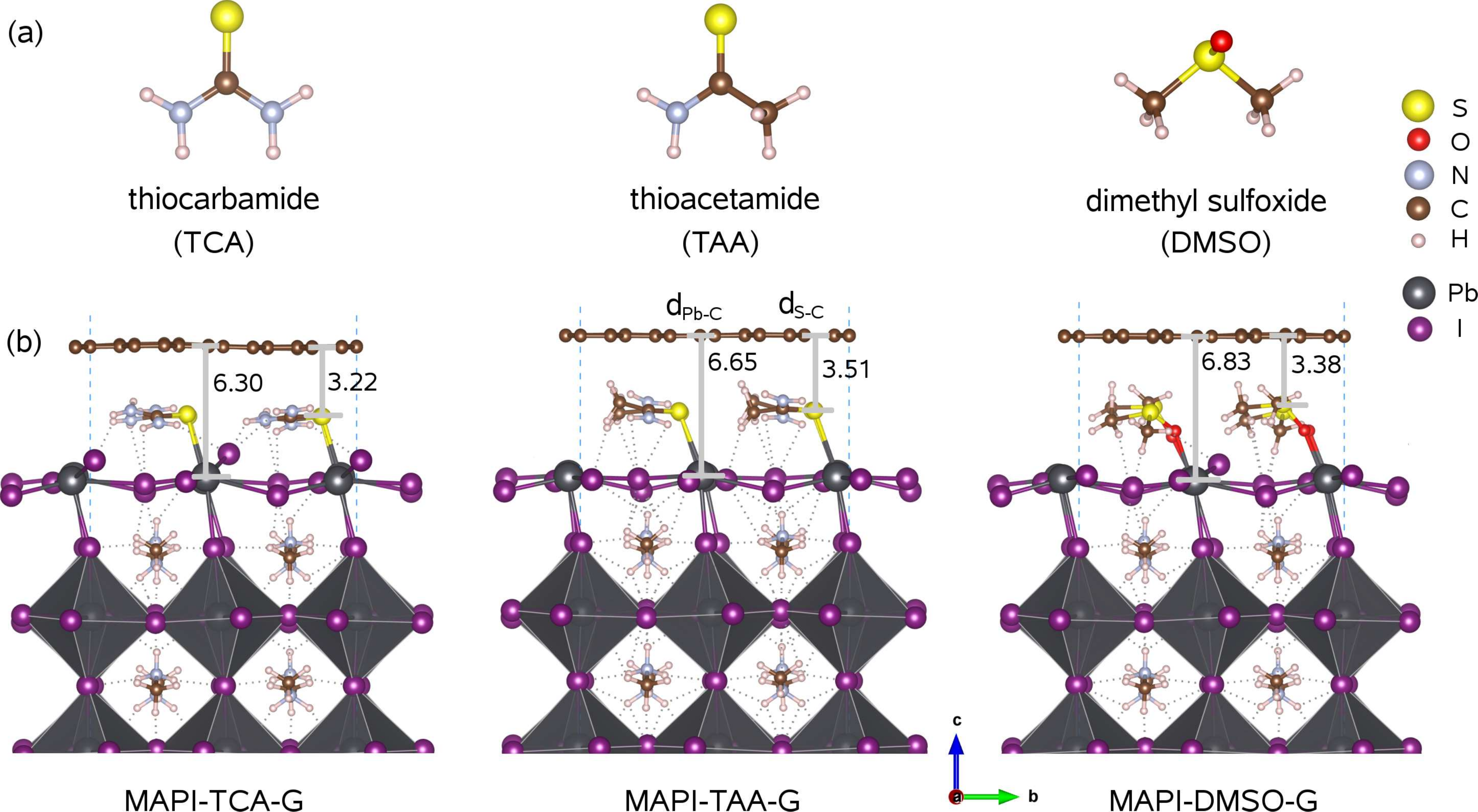} &
\includegraphics[clip=true,scale=0.4]{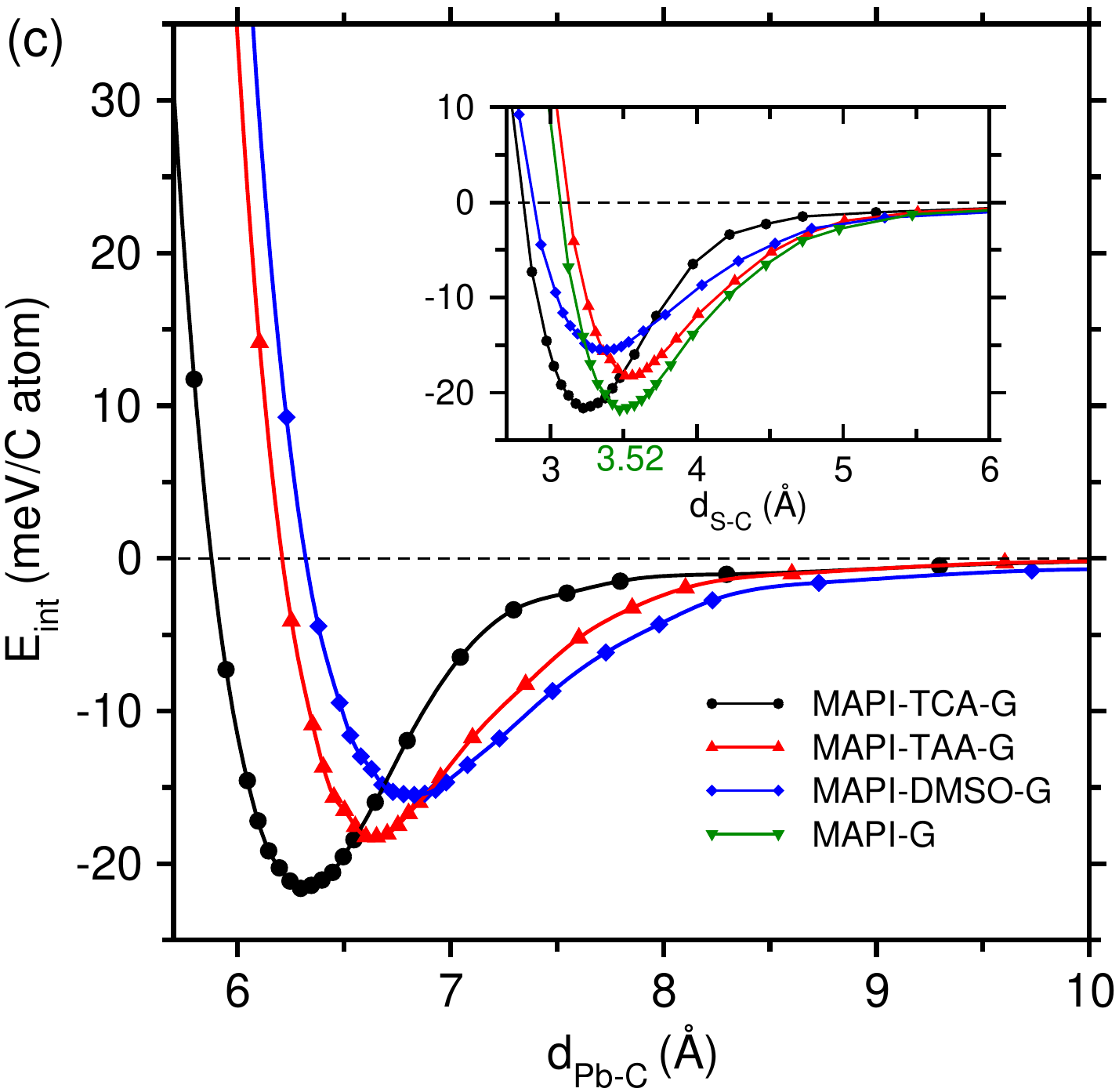}
\end{tabular}
\caption{(a) Molecular structures of the Lewis base molecules thiocarbamide (TCA), thioacetamide (TAA) and dimethyl sulfoxide (DMSO).
(b) Atomistic geometries of interfaces composed of MAPI(110) surface, molecular monolayer and graphene layer, optimized by using the pseudoatomic orbital basis sets of double-$\zeta$ (DZ) and BLYP + vdW functional. Hydrogen bonds between hydrogen and iodine atoms are denoted by dotted lines. The interlayer distances between the graphene sheet and the outermost Pb atom of the MAPI(110) substrate ($d_{\text{Pb-C}}$) and the outermost S atom of the Lewis base molecule ($d_{\text{S-C}}$) are indicated in \AA~unit. (c) The interlayer binding energy per carbon atom of graphene layer to the substrates, obtained by calculating the total energy difference of the interface system with the increasing interlayer distance from the system with infinity distance. Inset shows those in regard to the graphene-molecule interlayer distance.}
\label{fig1}
\end{figure*}

In this work, we devise MAPI interfaces in contact with monolayer of different donor bases, namely TCA, thioacetamide (TAA) and DMSO, and subsequently graphene layer.
These three molecules were considered in experiments to be typical Lewis bases that interact with MAPI and FAPI~\cite{Lee16acr,Ko19semsc}.
As shown in Figure~\ref{fig1}a, the main difference between these molecules consists in the electron-pair donor type and the functional groups bound to the central carbon or sulfur atom.
Given the experimental reports illustrating that incorporation of graphene and its derivatives into PSCs enhances the charge carrier transportation~\cite{Bonaccorso15s,Loh16jacs,Capasso16aami} and the graphite counter electrode (CE) replacing HTL and noble metal CE can improve the device stability~\cite{Ko19semsc,JLi17ra,LLiu15jacs,YRong14jpcl}, the graphene layer is also adopted here to cover the molecular layer.
We estimate the formation energies of various interface defects under different conditions (I-rich and Pb-rich) by density functional theory (DFT) calculations, using slab models (see SI for details).
Migrations of vacancy-mediated iodine atoms are examined to get an insight into stability improvement by forming adducts and interfaces with graphene.
To validate the theoretical findings, experiment has been performed.

We first explored the interface structures composed of a MAPI slab, a Lewis base monolayer, and a graphene layer.
Using the optimized lattice constants of $a=b=8.86$ \AA~and $c=12.38$ \AA~in the tetragonal phase, which are in agreement with the experimental values of 8.85 and 12.64 \AA~\cite{Poglitsch87jcp}, the (110) surface was created by cutting the crystal.
The (110) surface was accepted to be the most favorable nonpolar MAPI surface~\cite{Heo13np}, and moreover, known to minimize the lattice mismatch between MAPI and graphene structures~\cite{Zibouche18jpcc,Volonakis15jpcl}.
Given the fact that the \ce{PbI2} termination is the most stable non-defective (110) surface termination~\cite{Uratani17jpcl, Haruyama14jpcl, Haruyama16acr}, we made a slab model for the \ce{PbI2}-terminated (110) surface with the lattice constants of $a=12.36$ \AA~and $b=12.53$ \AA.
The slab thickness consists of five \ce{PbI2} layers and in-between four MAI layers (204 atoms), and vacuum thickness of over 30 \AA.
Surface relaxations on both the top and bottom surfaces were performed while keeping the atomic positions of middle layers fixed at the bulk positions.
On top of the relaxed surface, four Lewis base molecules at most were assumed to be adsorbed to form a monolayer, covering the MAPI surface that contains four MA groups in one layer of the slab.
After getting the rough adsorbate configurations on the relaxed surface by performing the Monte Carlo (MC) simulations, we carried out again atomic relaxations in the slabs for the molecular adsorbed MAPI(110) surfaces.
Finally, the graphene layer was modeled by a rectangular $(5\times3)$ supercell (60 atoms), which has the cell sizes matching well with the MAPI(110) surface~\cite{Zibouche18jpcc}, and was placed on the molecular monolayer.
Figure~\ref{fig1}b shows these interface structures optimized in this work (see Figure S3 for entire supercells).
\begin{table}[!th]
\caption{\label{tab1}Interlayer distance between the graphene layer and the outermost Pb atom of \ce{PbI2} layer of MAPI(110) surface $d_{\text{Pb-C}}$ and the outermost S atom of Lewis bases $d_{\text{S-C}}$ in \AA, formation energy $E_{\text{f}}$ in kJ/mol, binding energy per surface area in eV/\AA$^2$, interlayer binding energy per carbon atom of graphene to the surface $E_{\text{int}}$ in meV/atom, interface dipole moment $p$ in Debye unit, and work function change $\Delta \phi$ in eV.}
\small
\begin{tabular}{l@{\hspace{6pt}}c@{\hspace{6pt}}c@{\hspace{6pt}}c@{\hspace{6pt}}c@{\hspace{6pt}}c@{\hspace{6pt}}c@{\hspace{6pt}}c}
\hline
system & $d_{\text{Pb-C}}$ & $d_{\text{S-C}}$ & $E_{\text{f}}$ & $E_{\text{b}}$ & $E_{\text{int}}$ & $p$ & $\Delta\phi$ \\
\hline
MAPI-G      & --   & 3.52 & $-$188 & $-$12.58 & $-$20.74 &  --  & 0.37  \\
MAPI-TCA-G  & 6.30 & 3.22 & $-$561 & $-$13.82 & $-$21.61 & 4.16 & 0.95  \\
MAPI-TAA-G  & 6.65 & 3.51 & $-$390 & $-$11.73 & $-$18.23 & 3.74 & 1.15  \\
MAPI-DMSO-G & 6.83 & 3.38 &~~$-$66 &~~$-$9.52 & $-$15.30 & 4.37 & 1.69  \\
\hline
\end{tabular}
\end{table}
\begin{figure*}[!th]
\centering
\begin{tabular}{lll}
\hspace{1.7pt}\includegraphics[clip=true,scale=0.078]{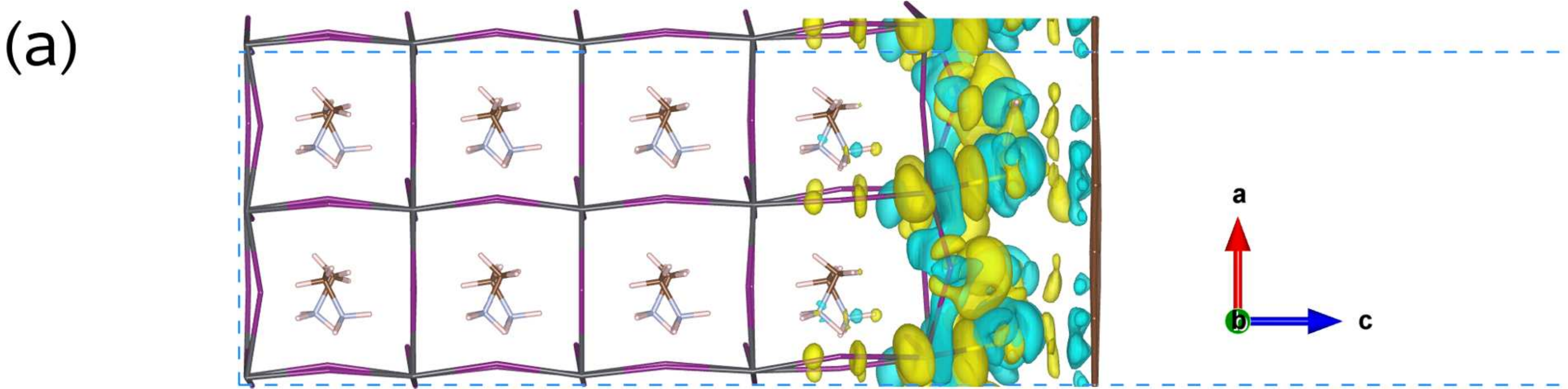}&
\hspace{13.8pt}\includegraphics[clip=true,scale=0.077]{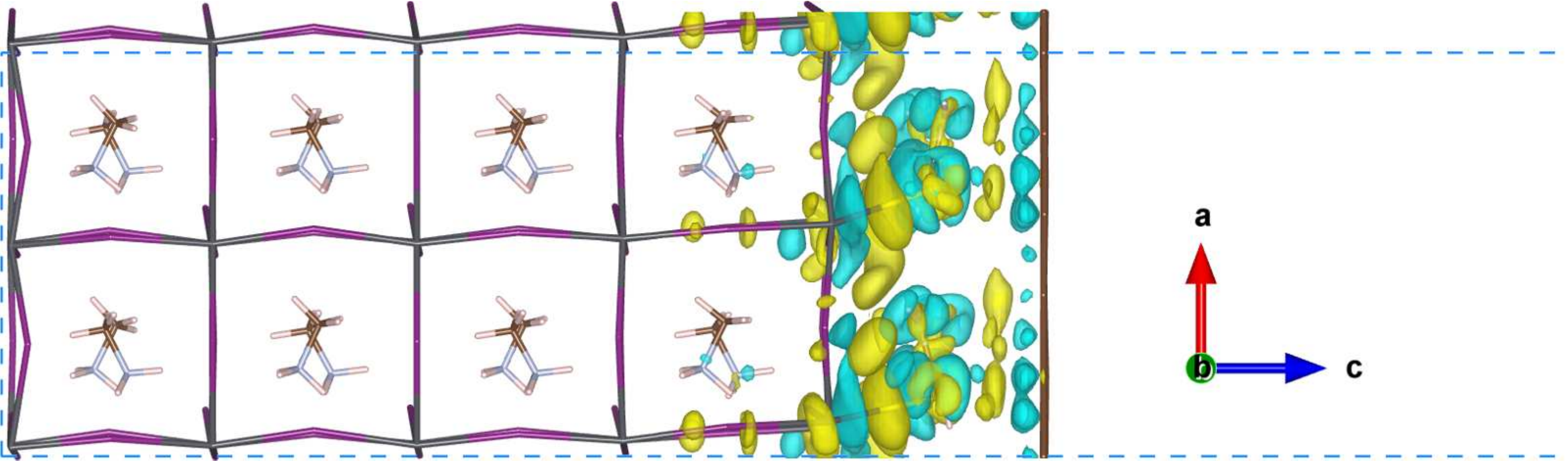}&
\hspace{13.8pt}\includegraphics[clip=true,scale=0.078]{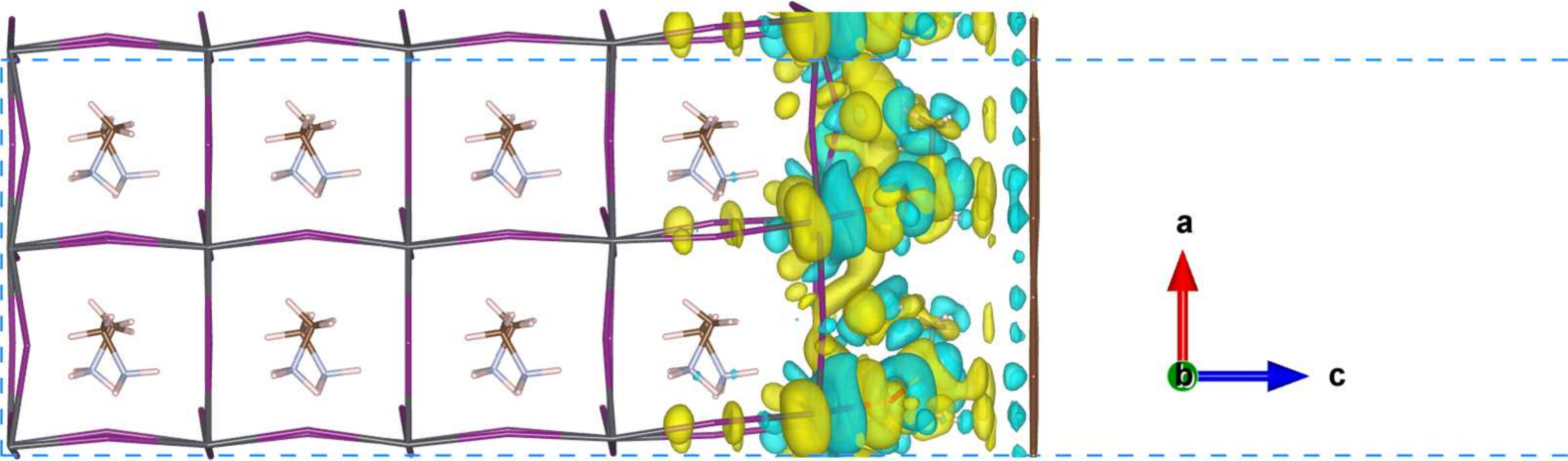} \\
\includegraphics[clip=true,scale=0.38]{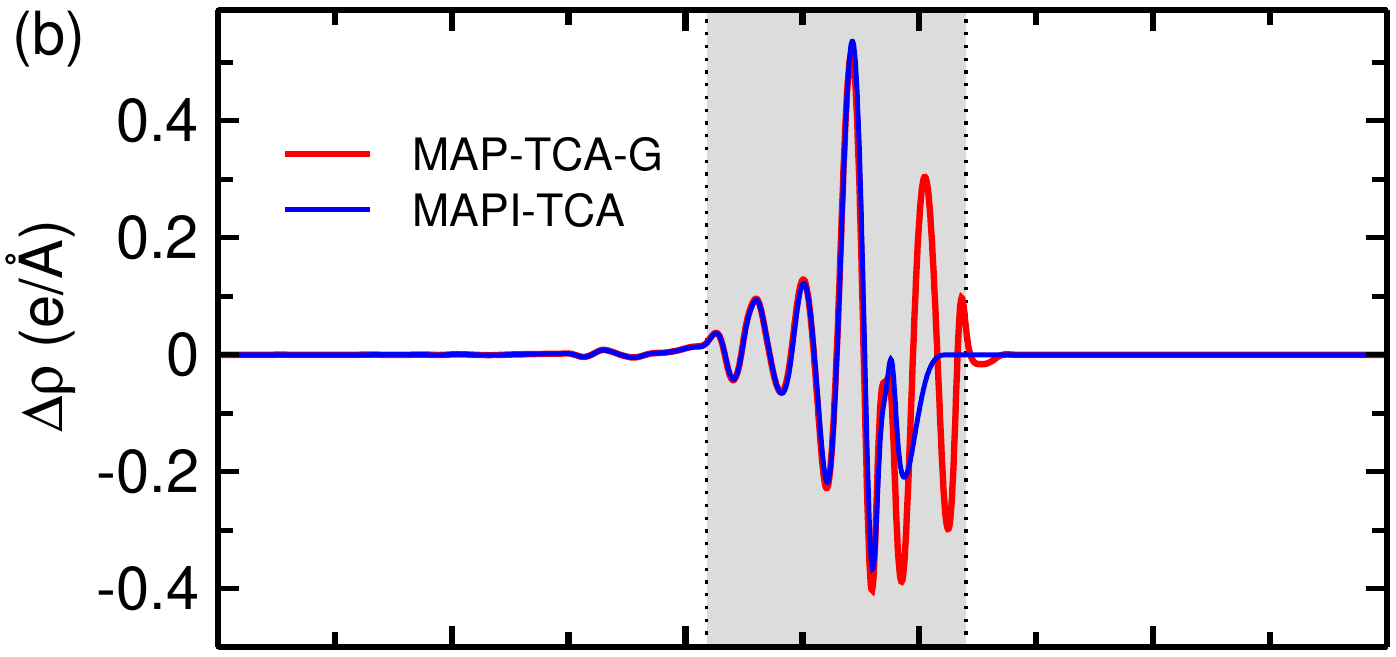} &
\includegraphics[clip=true,scale=0.38]{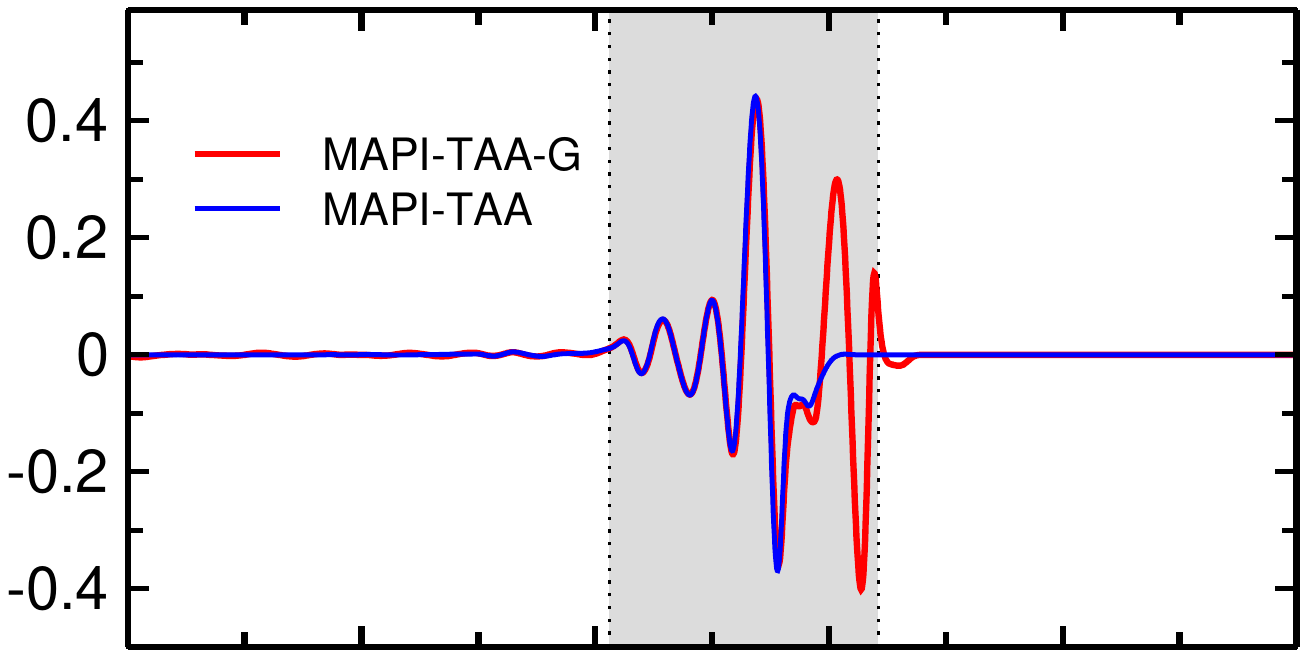} &
\includegraphics[clip=true,scale=0.38]{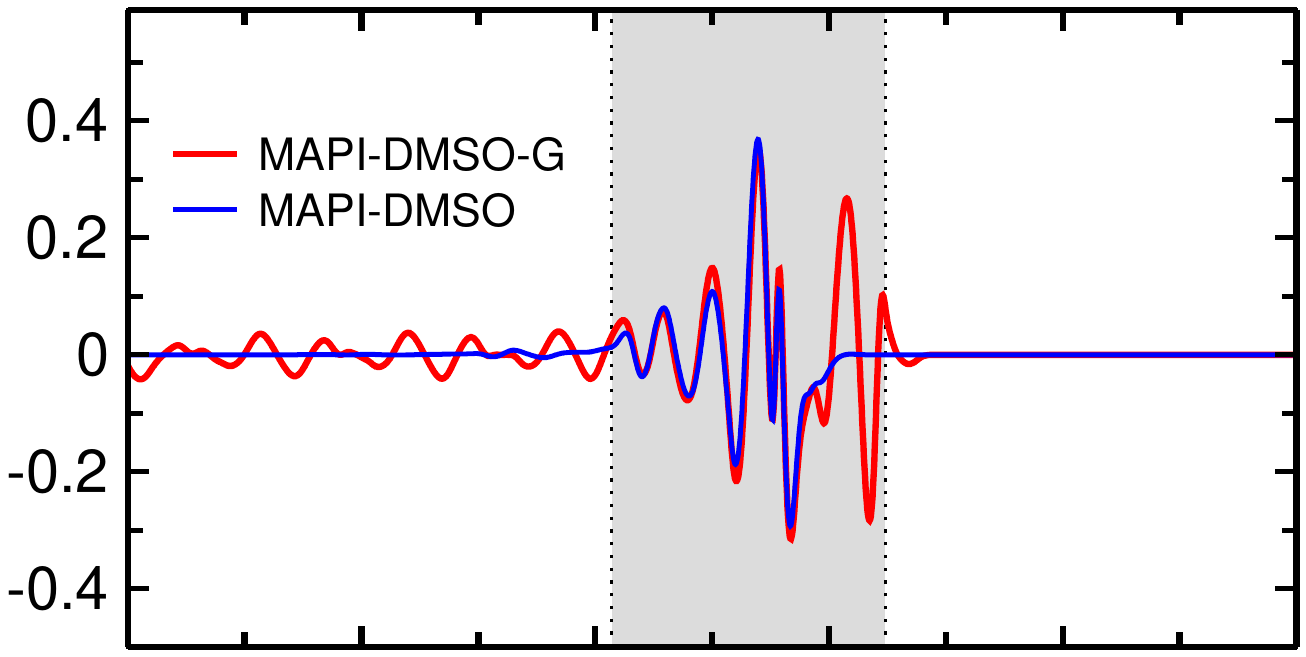} \\
\hspace{1.5pt}\includegraphics[clip=true,scale=0.38]{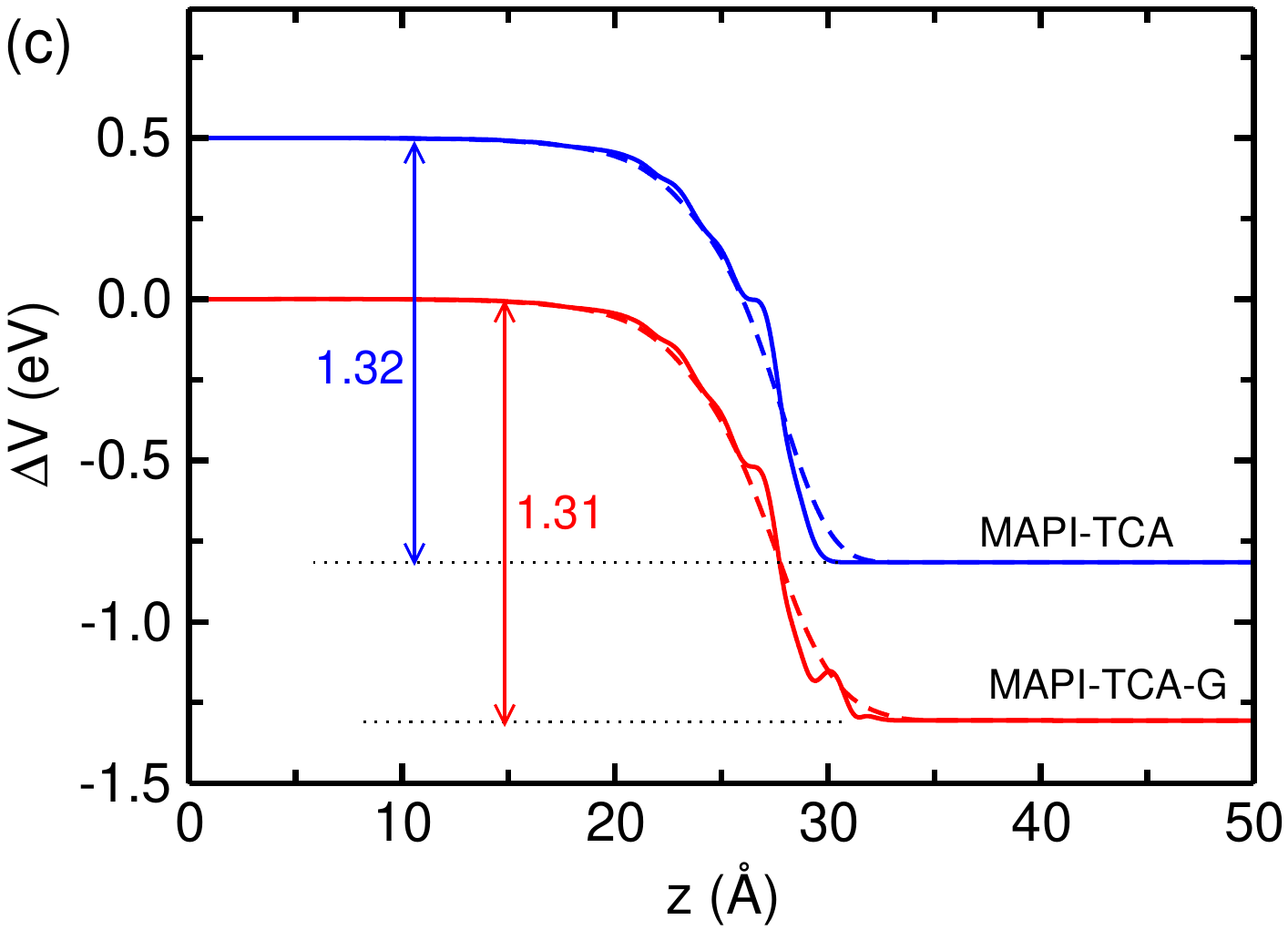} &
\includegraphics[clip=true,scale=0.38]{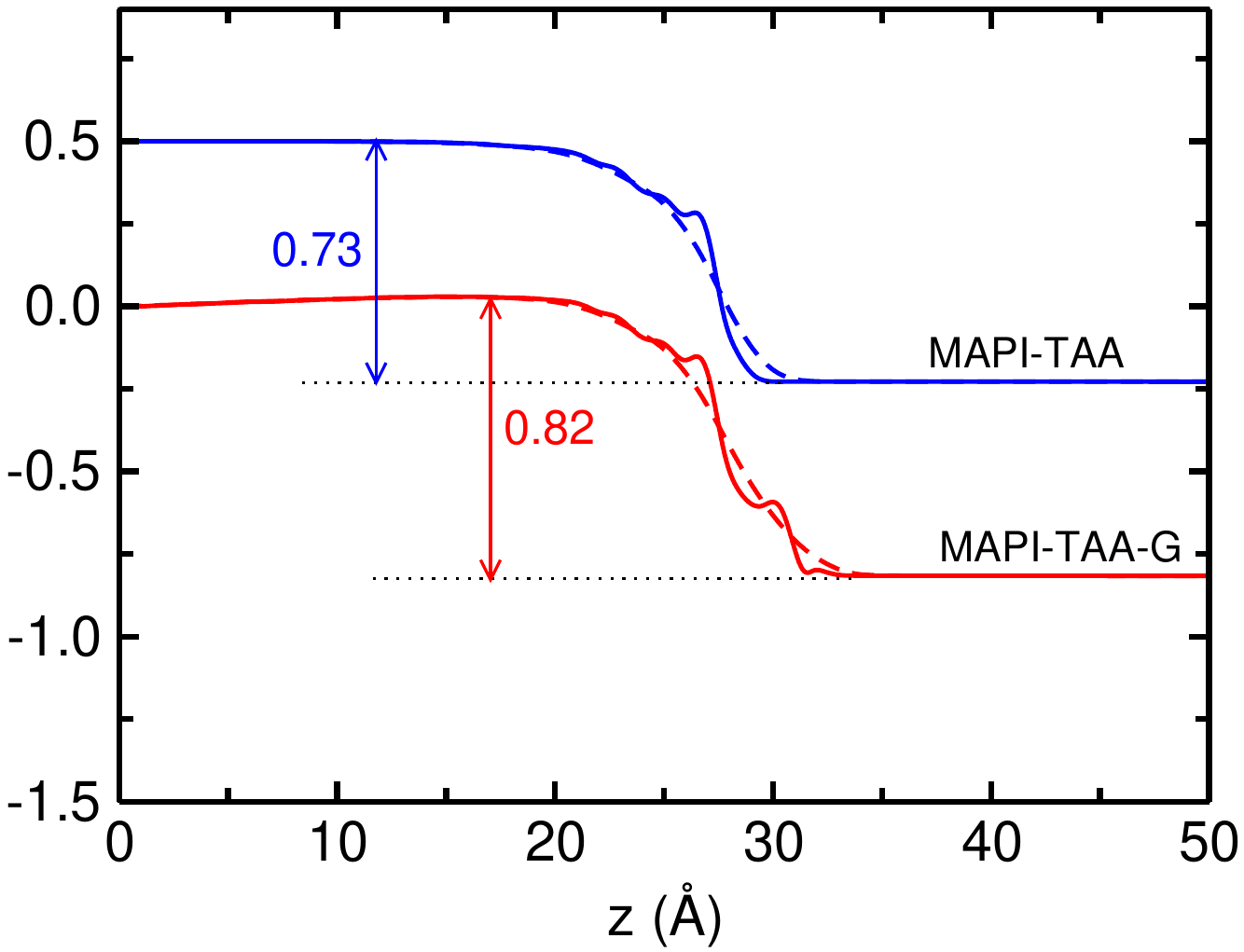} &
\includegraphics[clip=true,scale=0.38]{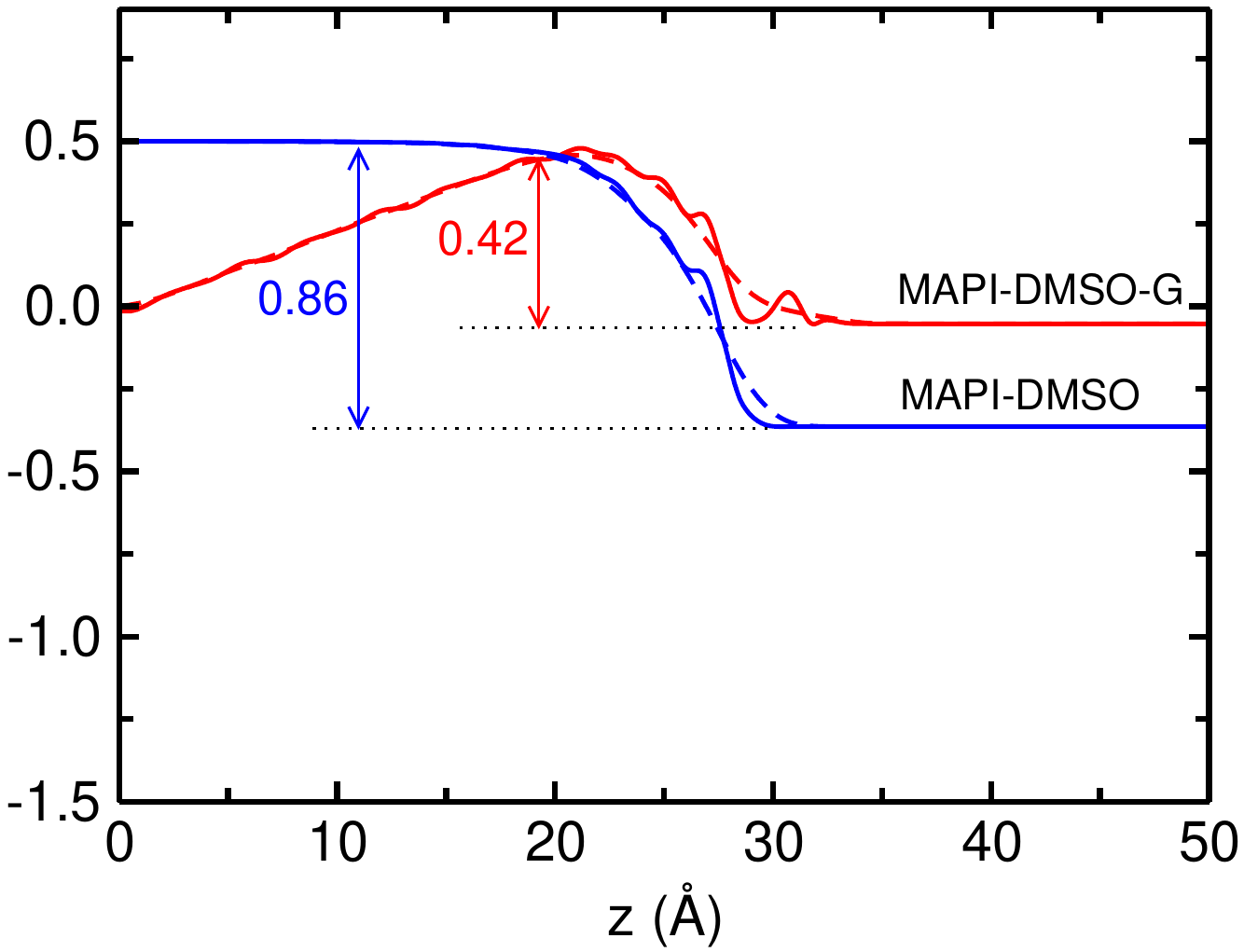} \\
\end{tabular}
\caption{(a) Isosurface view of spatial electron density difference at the same value of 0.005 $|e|$/\AA$^3$ upon the formation of MAPI-mol-G (mol: TCA, TAA, DMSO) interface; the yellow and cyan color indicate the electron gain and loss respectively. (b) Planar average electron density difference ($\Delta\rho$) and (c) electrostatic potential difference ($\Delta V$) integrated over the $x-y$ plane for the MAPI-mol (blue line) and MAPI-mol-G (red line) systems as a function of $z$ coordinate (perpendicular to the surface). Grey-color indicates the interface regions, the dashed lines the macroscopic average, and the potential drop values are denoted in eV unit.}
\label{fig2}
\end{figure*}

Regarding energetics, the adsorption energies per molecule for the cases of four molecules adsorption were found to be lower than the cases of one or two molecules adsorption (see Table S1 in SI).
For the MAPI-mol-G, the formation energies were estimated to be negative as $-561$, $-390$, and $-66$ kJ/mol for the TCA, TAA and DMSO systems respectively (Table~\ref{tab1}), indicating that, while the interfaces are likely to form exothermically from MAPI, Lewis base molecules and graphene, the formability can be remarkably enhanced from the $O$-donor DMSO to the $S$-donors TAA and TCA molecules.
We then calculated the binding energies per surface area $E_{\text{b}}$ in these interfaces, as $-$13.8, $-$11.7 and $-$9.5 meV/\AA$^2$.
For reference, the binding energy of MAPI-G interface was also calculated as $-$12.6 meV/\AA$^2$, which is slightly lower than the previous theoretical value of $-$8 meV/\AA$^2$~\cite{Zibouche18jpcc}.
As a measure of binding strength between the graphene layer and the MAPI-mol substrates, the interlayer binding energies $E_{\text{int}}$ were evaluated as a function of interlayer distance (Figure~\ref{fig1}c).
Going from the DMSO to the TAA and to the TCA systems, $E_{\text{int}}$ was found to increase from $-$15.3 to $-$18.2 and to $-$21.6 meV, and accordingly the equilibrium interlayer distance $d_{\text{Pb-C}}$ to decrease from 6.83 to 6.65 and to 6.30 \AA.
Note that for the MAPI-G interface $d_{\text{Pb-C}}$ distance was found to be 3.52 \AA~in good agreement with the previous value of 3.54 \AA~\cite{Zibouche18jpcc}.
Again, these binding informations revealed a gradual enhancement of interactions between MAPI, molecule and graphene from DMSO to TAA and to the TCA molecules.
It is worth noting that the binding energies ($E_{\text{f}}$ and $E_{\text{int}}$) of the MAPI-TCA-G interface ($-$13.8 meV/\AA$^2$~and $-$21.6 meV) are slightly higher than those of MAPI-G interface ($-$12.6 and $-$20.4), indicating a certain role of the TCA molecule in enhancing the contact between MAPI and graphene.
The lone-pair electrons donors S or O atoms of Lewis base were observed to be bound with the electron-pair acceptors Pb atoms of MAPI, while the $-$\ce{NH2} and $-$\ce{CH3} functional groups of the Lewis bases can form the hydrogen bonds with the I atoms of MAPI.
Since the $S$-donor was known to be stronger than the $O$-donor~\cite{Lee16acr,Wharf76cjc} and the $-$\ce{NH2} group is more favorable for hydrogen bond than the $-$\ce{CH3} group, the TCA molecule containing a S atom and two \ce{NH2} groups should have the highest binding strength while the DMSO base containing a O atom and two \ce{CH3} groups has the lowest strength.

Figure~\ref{fig2}a and b show the electron density difference for the three interface systems.
For the same isosurface value of 0.005 $e$/\AA$^3$, the charge redistribution becomes weaker going from TCA to DMSO.
The charge accumulation is seen near the outermost Pb atoms while the depletion around the S or O atoms of Lewis base, and when adding the graphene layer, additional charge depletion appears near the layer, indicating that the molecules and graphene donate electrons to the MAPI.
Such electron charge redistribution induces an interface dipole moment $p$, which helps to move the photo-generated electrons from the perovskite to the graphene layer.
The interface dipole can be divided into two terms: the bond dipole caused by the electron rearrangement and the molecular dipole induced by the deformation of the molecular structure~\cite{Liu16jpcl}.
The molecular dipole of DMSO was found to be the highest as 4.1 D, while TCA has the lowest value in the opposite direction of $-$1.2 D.
Inversely, the bond dipole increases from DMSO (0.2 D) to TCA (5.4 D), yielding the somewhat similar interface dipoles of 4.2, 3.7 and 4.4 D for the TCA, TAA and DMSO systems respectively. (for details, see SI).
The created interface dipole causes also the electrostatic potential drops as shown in Figure~\ref{fig2}c, leading to the downward surface band bending with the formation of charge accumulation layer at the surface and the work function reduction of the bare MAPI(110) surface~\cite{Stahler17cp}.
The work function decreases were found to be 1.31, 0.82 and 0.42 eV from the bond dipole (Figure~\ref{fig2}c) while $-$0.38, 0.32 and 1.24 eV from the molecular dipole, yielding the total reductions of 0.95, 1.15, and 1.69 eV upon the formation of the TCA, TAA, and DMSO interfaces.
When a photon is absorbed in the MAPI layer, an electron is excited from its valence band maximum (VBM) to the conduction band minimum (CBM), and attracted by the interface dipole toward graphene (see Figure S7).

\begin{figure*}[!th]
\centering
\begin{tabular}{c@{}cc}
\includegraphics[clip=true,scale=0.35]{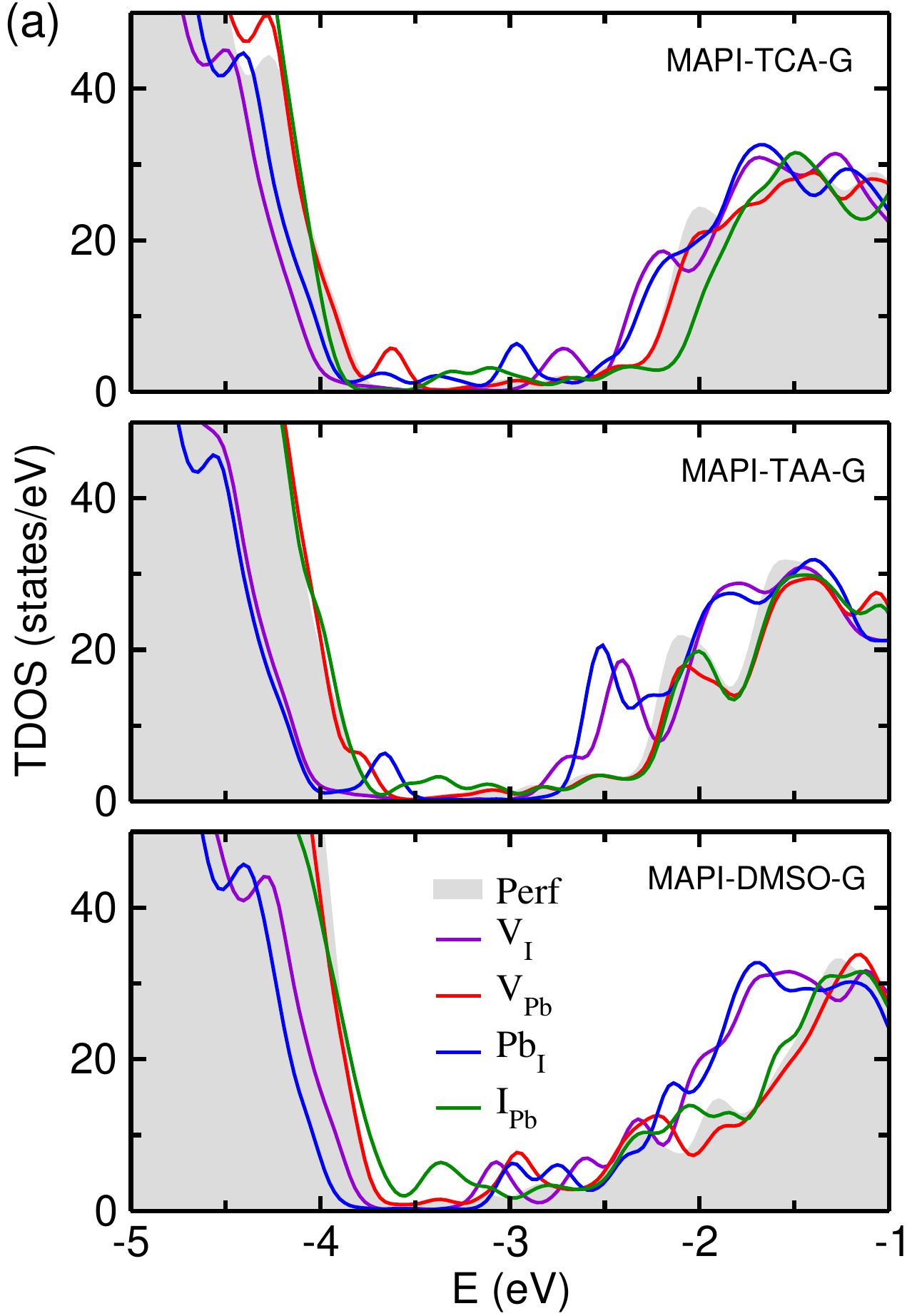} & 
\includegraphics[clip=true,scale=0.34]{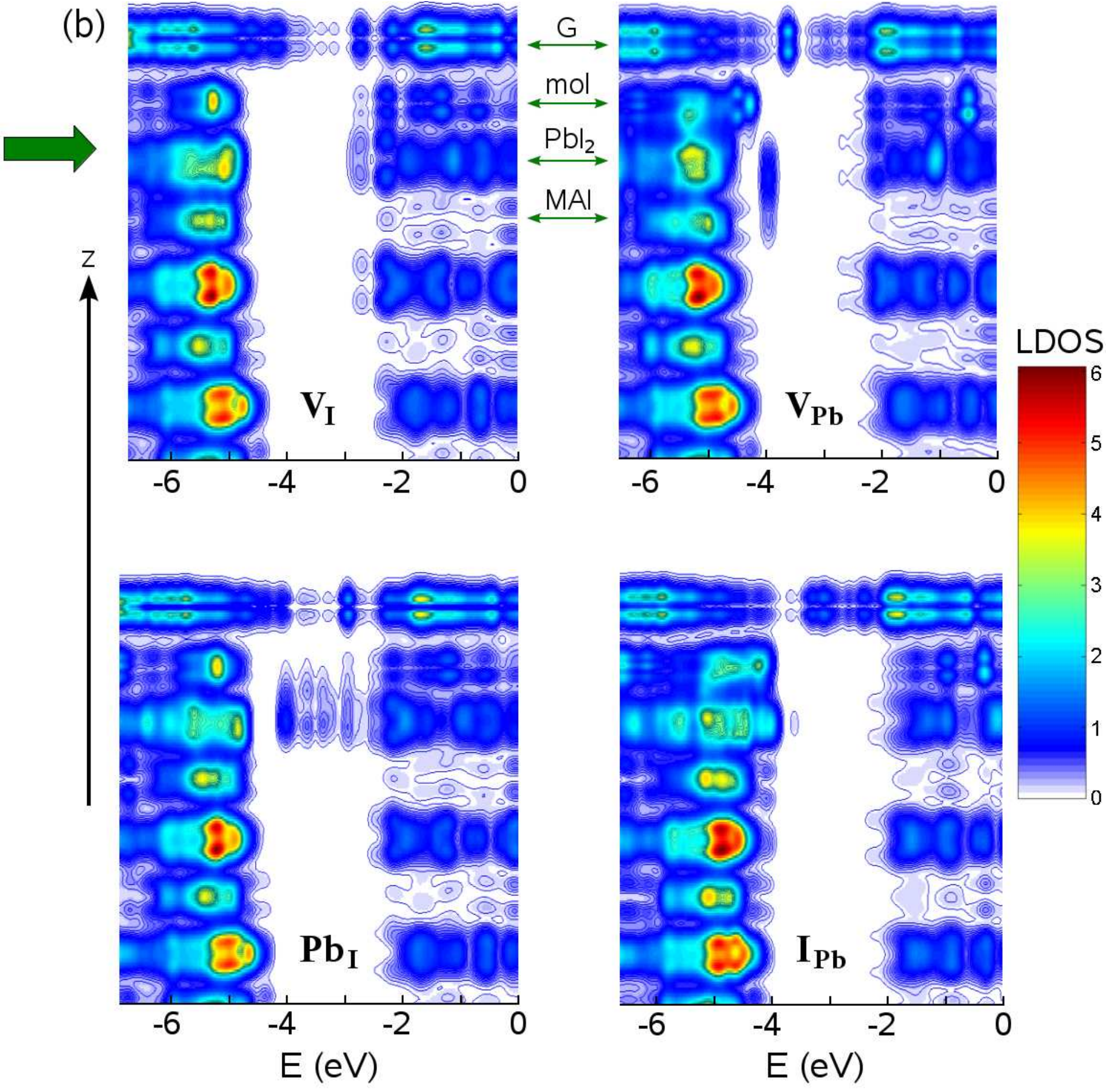} &
\includegraphics[clip=true,scale=0.062]{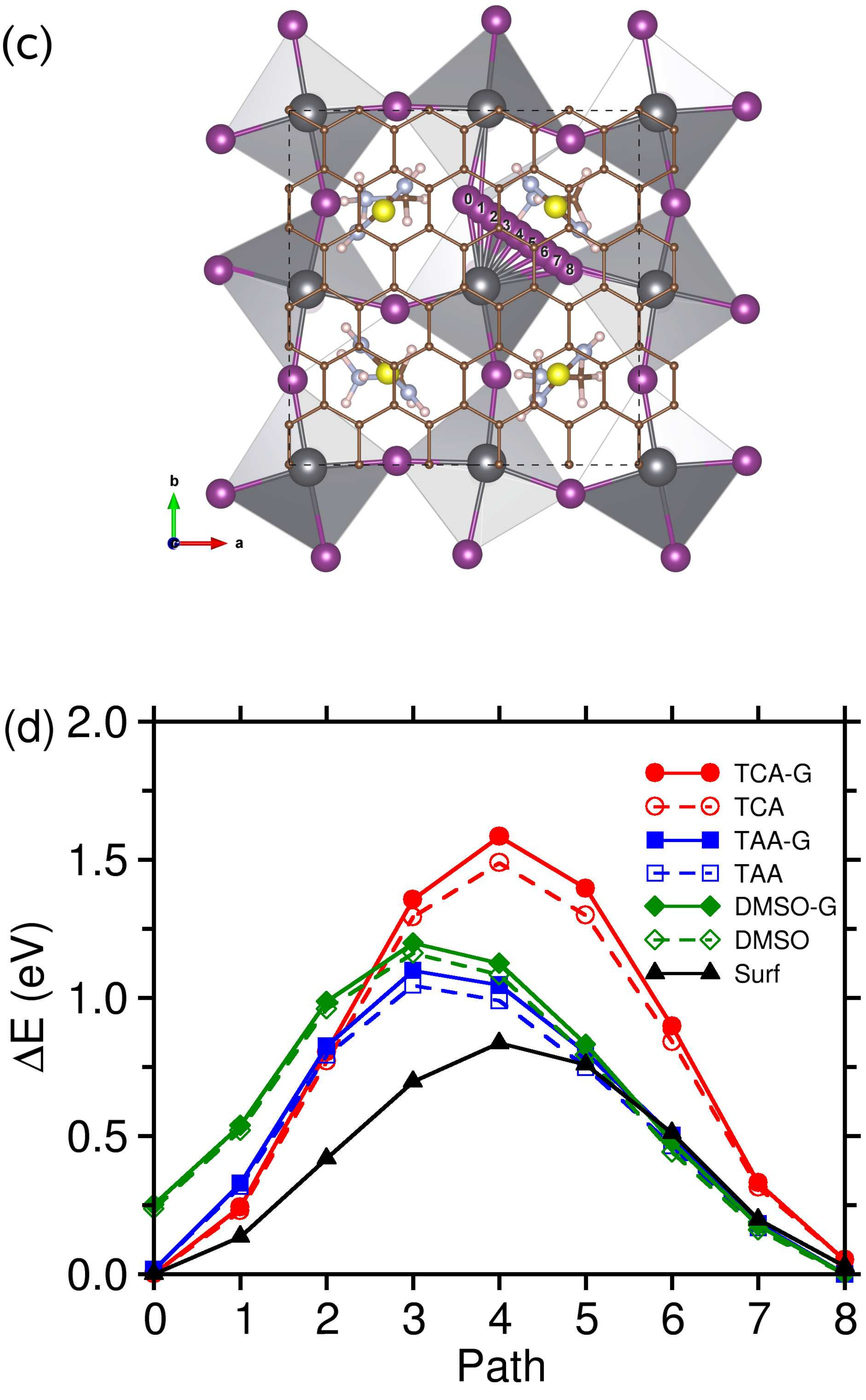} \\
\end{tabular}
\caption{(a) Total density of states (TDOS) in MAPI-mol-G (mol: TCA, TAA, DMSO) interfaces without and with the I- and Pb-related point defects such as vacancies (V$_{\text{I}}$, V$_{\text{Pb}}$) and antisites (Pb$_{\text{I}}$, I$_{\text{Pb}}$). (b) Local DOS (LDOS) in the TCA interface with the defects. (c) Top view of vacancy-mediated I atom migration along the edge of \ce{PbI5} octahedron on the MAPI(110) surface, and (c) their activation barriers in the bare surface, MAPI-mol and MAPI-mol-G interface systems.}
\label{fig3}
\end{figure*}
\begin{table}[!th]
\caption{\label{tab2}Formation energy of neutral point defects associated with surface Pb and I atoms, i.e., vacancies (V$_\text{I}$, V$_\text{Pb}$) and antisites (Pb$_\text{I}$, I$_\text{Pb}$), under the I-rich (or Pb-poor) and Pb-rich (or I-poor) conditions. Values are given in eV unit.}
\small
\begin{tabular}{lccc@{\hspace{5pt}}cccc}
\hline
defect & \multicolumn{3}{c}{I-rich (Pb-poor)} & & \multicolumn{3}{c}{Pb-rich (I-poor)}\\
\cline{2-4} \cline{6-8}
       & TCA & TAA & DMSO && TCA & TAA & DMSO \\
\hline
V$_\text{I}$  & 2.24 & 1.85 & 1.90 && 0.99 & 0.59 & ~~0.65 \\
V$_\text{Pb}$ & 0.76 & 0.81 & 1.10 && 3.26 & 3.31 & ~~3.60 \\
Pb$_\text{I}$ & 5.15 & 4.51 & 3.20 && 1.39 & 0.76 &$-$0.56 \\
I$_\text{Pb}$ & 0.67 & 0.87 & 0.99 && 4.43 & 4.63 & ~~4.75 \\
\hline
\end{tabular}
\end{table}
We then turn our attention to the surface defects to see whether the surface trap passivation is likely to occur in these interfaces.
We calculated the formation energies of vacancies (V$_\text{I}$, V$_\text{Pb}$) and antisites (Pb$_\text{I}$, I$_\text{Pb}$) under the I-rich and Pb-rich conditions, as presented in Table~\ref{tab2}.
Under the I-rich condition, the dominat defects with lower formation energies, which were found to be different depending on the condition~\cite{Kye18jpcl,MHDu15jpcl,Buin14nl,Yin114apl,JKim14jpcl}, were formed on the Pb position, i.e., the Pb atom vacancies (V$_\text{Pb}$) and the antisite I atoms (I$_\text{Pb}$).
Meanwhile, the I atom vacancies (V$_\text{I}$) and the antisite Pb atoms (Pb$_\text{I}$) have relatively lower formation energies under the Pb-rich condition.
It was found that the formation energies of the I-related defects decrease going from the TCA to the DMSO system, while those of the Pb-related defects increase.
Regarding the defect formation, therefore, the I-rich condition is more desirable for the TCA system while the Pb-rich condition for the DMSO system.
As shown in Figure~\ref{fig3}a for the total density of states (DOS) in the interfaces, for the TCA and TAA systems, only the Pb$_\text{I}$ defects were revealed to create the deep trap states near the middle point between VBM and CBM, while the other defects induce the shallow transition states.
However, due to the high formation energies under the I-rich condition, the Pb$_\text{I}$ defects are not likely to be formed under this condition.
Hence, the I-rich condition is favorable for surface trap passivation in the TCA and TAA systems.
On the contrary, we see deep trap states for all the relevant defects in the DMSO system, indicating a poor surface trap passivation in the MAPI-DMSO-G system.
Figure~\ref{fig3}b shows the local DOS (LDOS) in the MAPI-TCA-G interface with the defects, indicating that only the Pb$_{\text{I}}$ defect exhibits deep trap states and thus the interface formation with TCA and graphene can suppress the surface trap states (for others, see Figures S10 and S11).
As a measure of stability of PSC~\cite{yucj17jmca, Aristidou17nc, Eames15nc, Azpiroz15ees, Haruyama15jacs}, we evaluated the activation barriers for vacancy-mediated I atom migration along the edge path of \ce{PbI5} octahedron on the surface (see Figure~\ref{fig3}c).
As shown in Figure~\ref{fig3}d, the barriers become higher when adsorbing the Lewis base molecules on the MAPI(110) surface, and further higher slightly upon covering with the graphene layer, indicating the stability enhancement by the interface formation.
Among the three molecules, TCA was found to raise the barrier for the highest, while TAA and DMSO have similar effects.
To sum up so far, the MAPI-TCA-G interface system, among three different systems, is the most desirable for enhancing the solar cell performance regarding the efficiency and stability.

\begin{figure*}[!th]
\centering
\begin{tabular}{rcc}
\includegraphics[clip=true,scale=0.6]{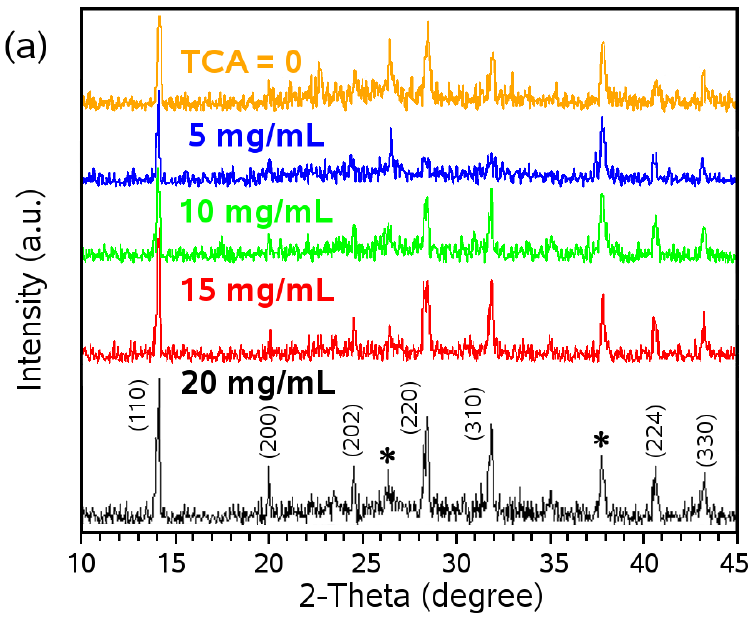} & 
\includegraphics[clip=true,scale=0.39]{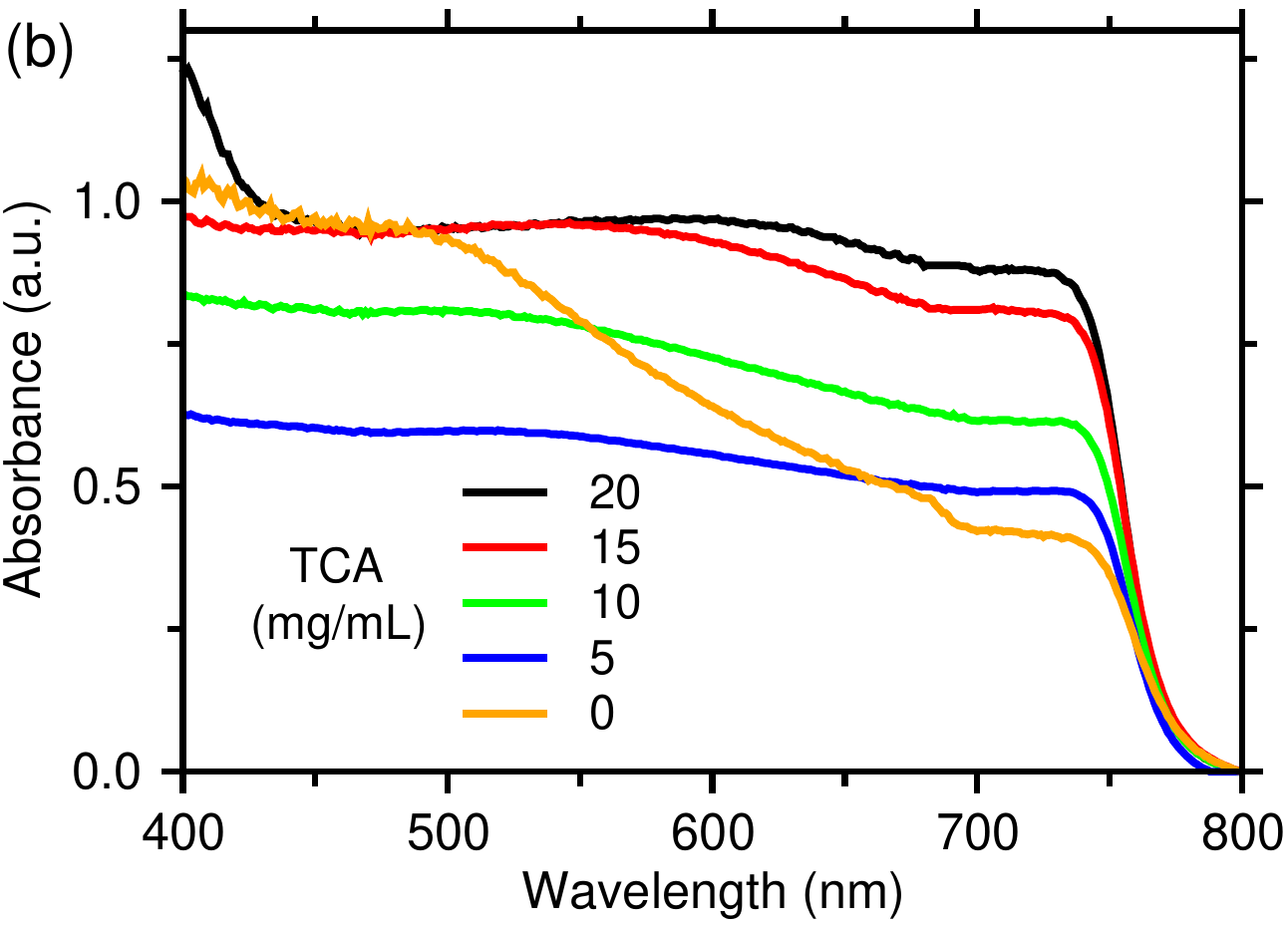}  &
\includegraphics[clip=true,scale=0.35]{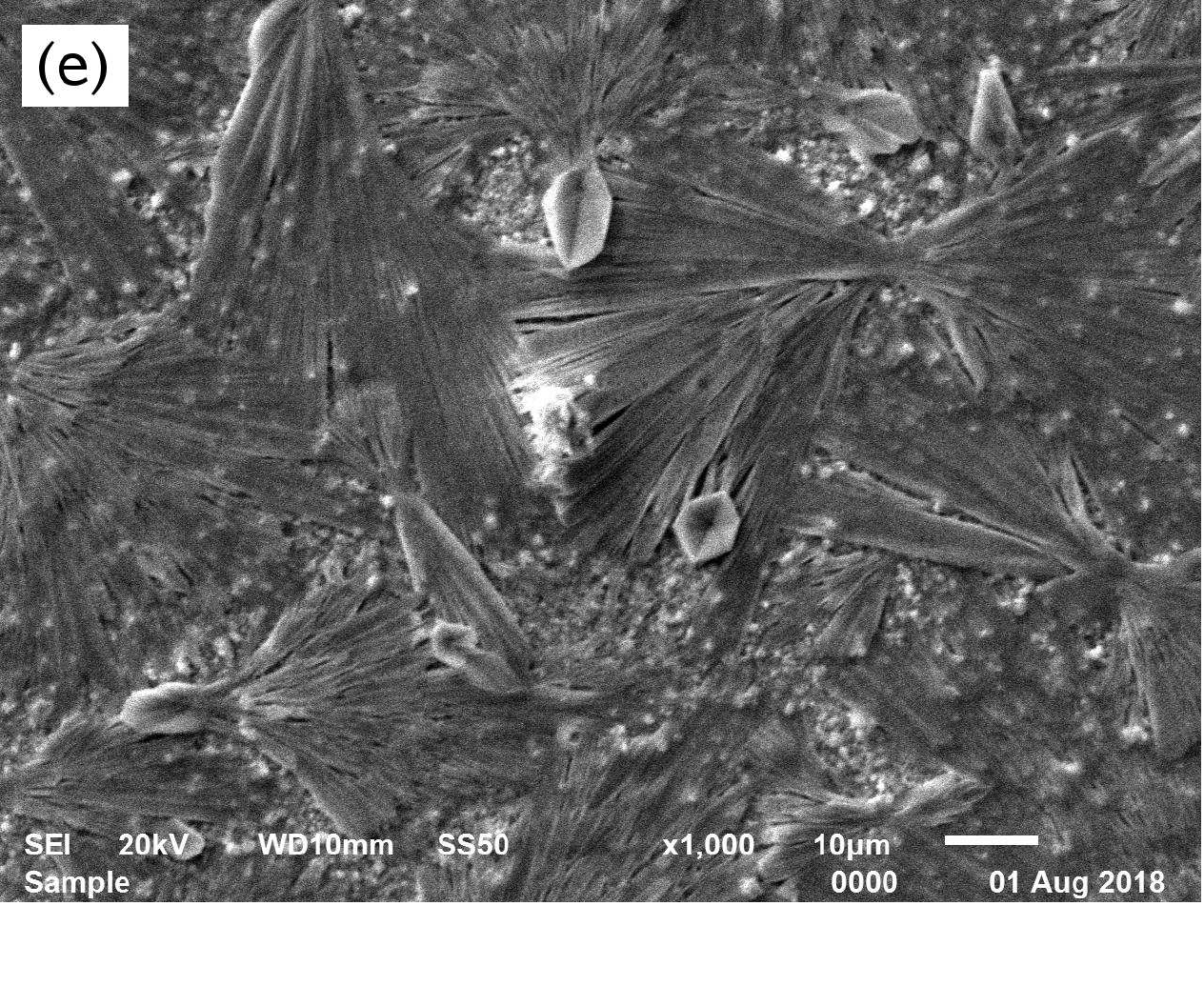} \\
\includegraphics[clip=true,scale=0.07]{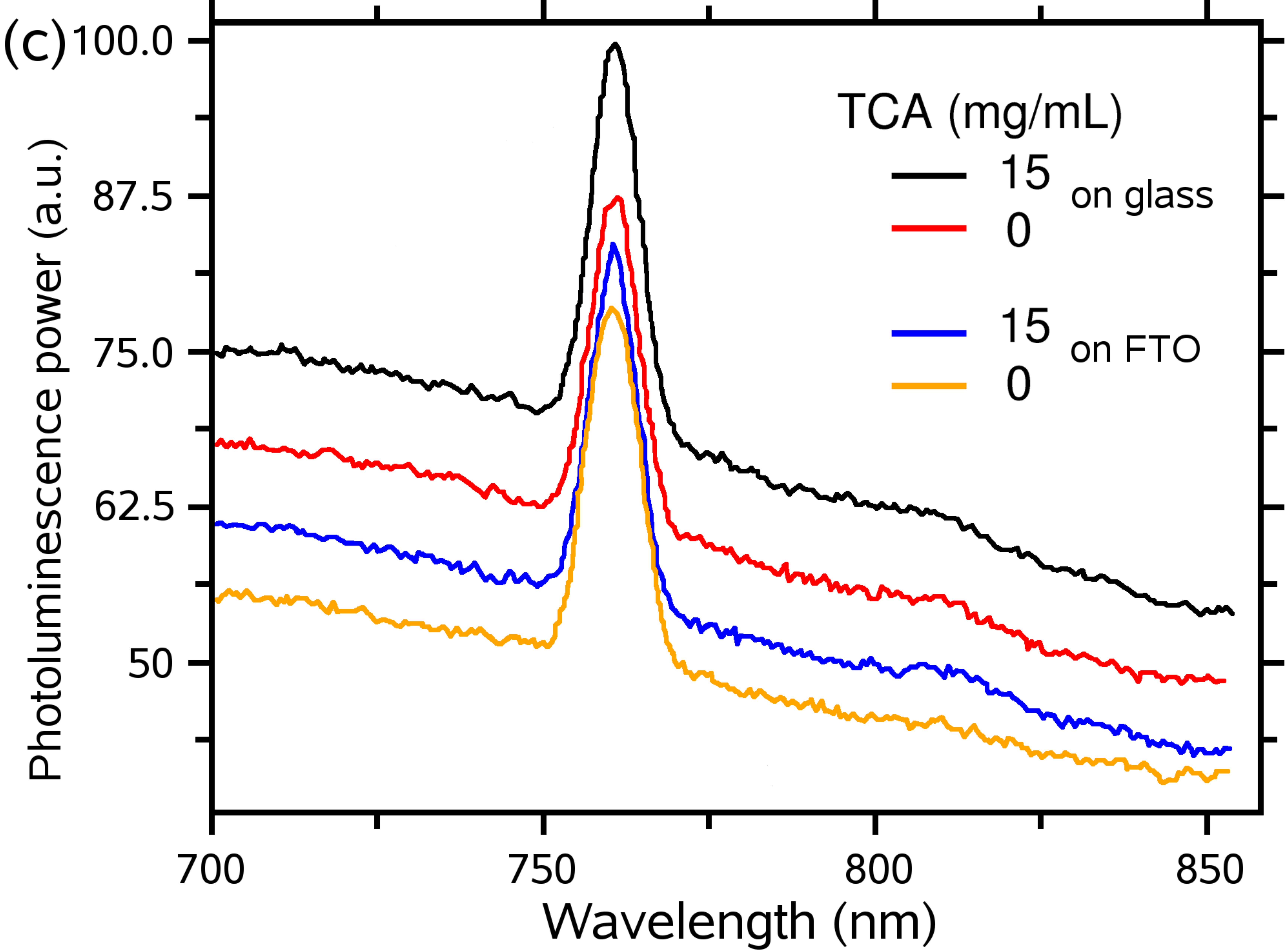} & 
\includegraphics[clip=true,scale=0.39]{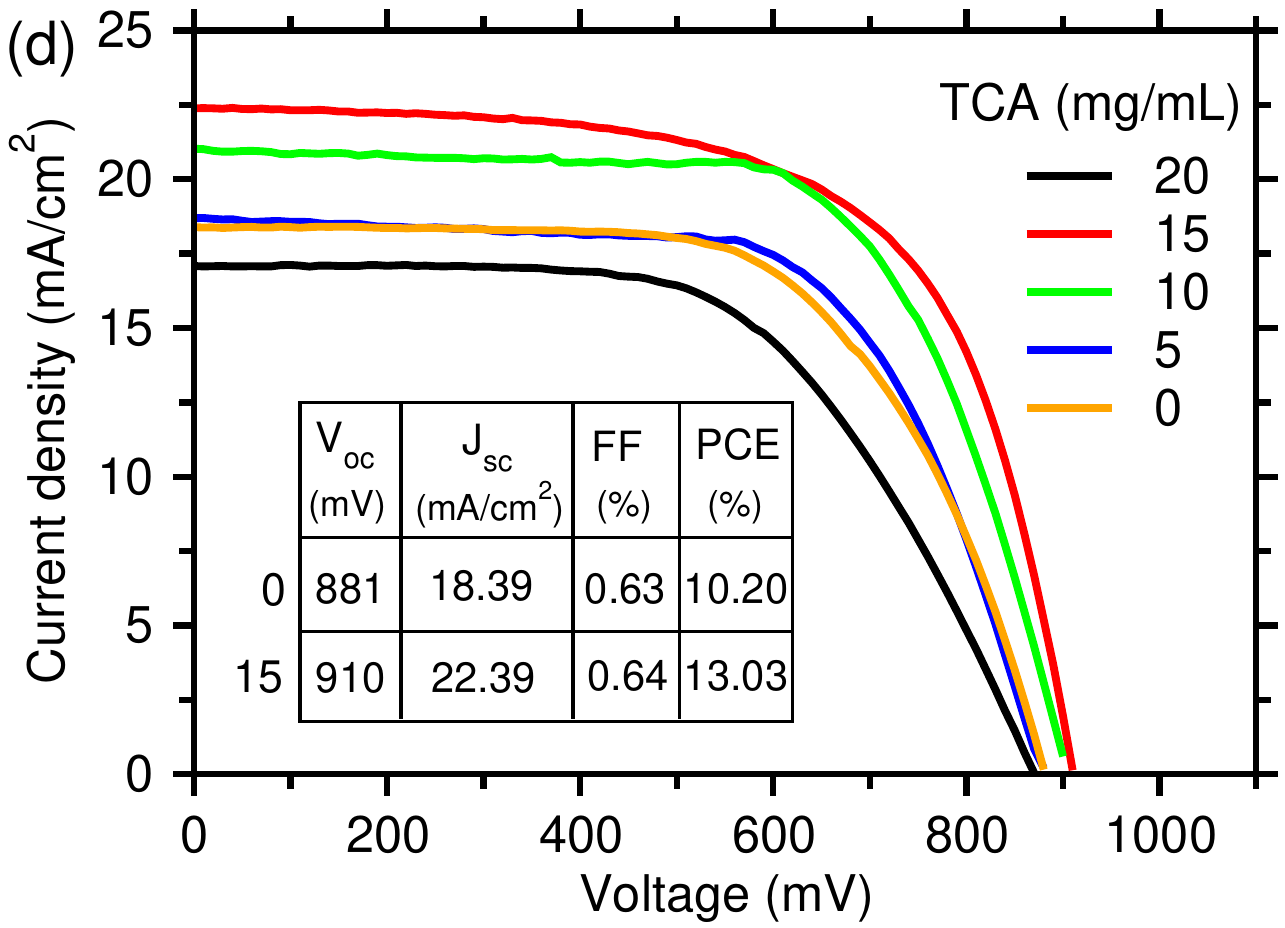}  &
\includegraphics[clip=true,scale=0.35]{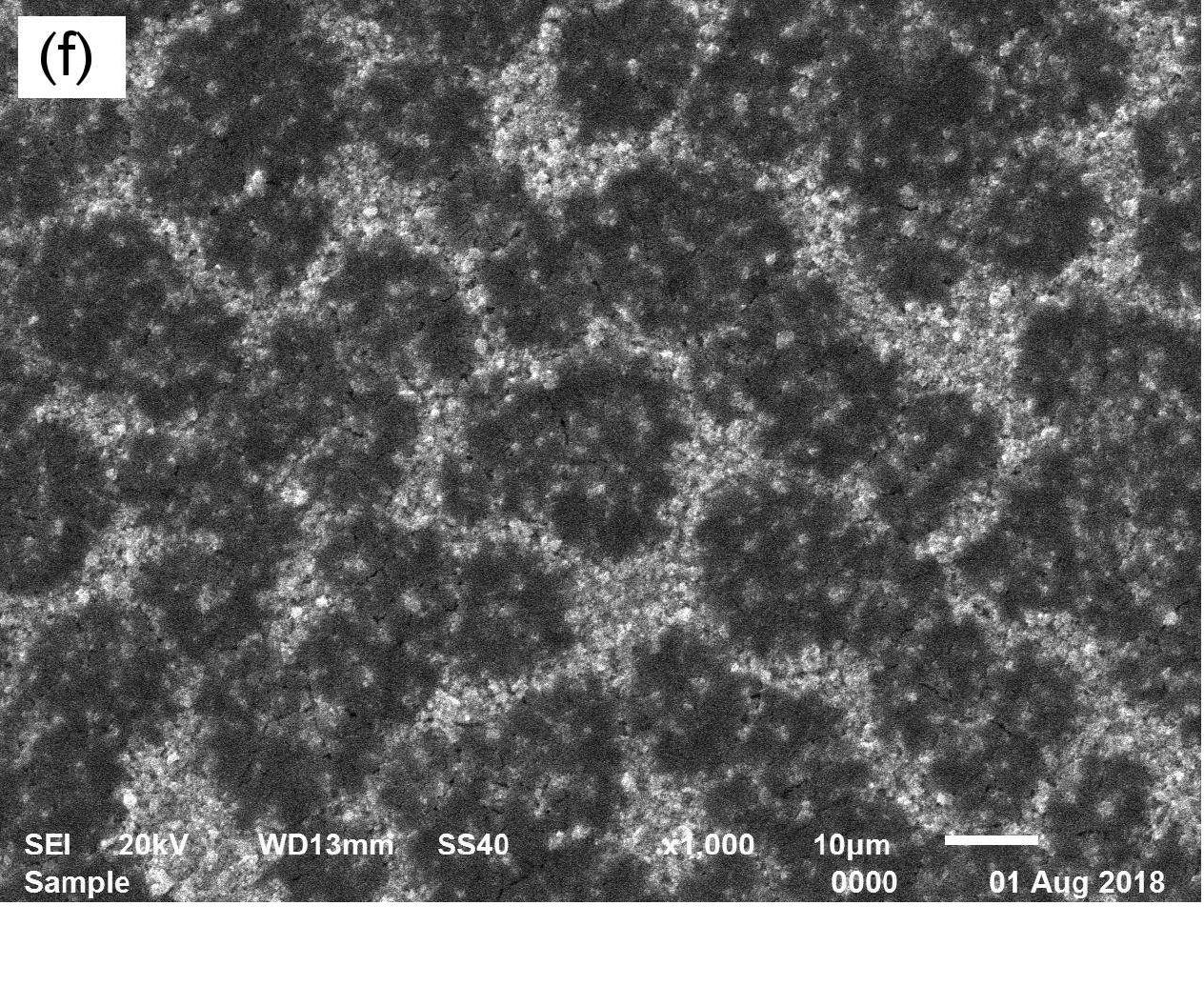} \\
\end{tabular}
\caption{Experimental investigation of the MAPI perovskites with TCA of various concentrations (0, 5, 10, 15, 20 mg/mL) and the solar cells. (a) XRD patterns for the perovskite films, where the asterisks correspond to the fluorine-doped tin oxide (FTO) as substrate, (b) absorption spectra, (c) photoluminescence spectral power and (d) photocurrent density versus voltage. SEM images for the perovskite films coated on FTO (d) without and (f) with 15 mg/mL TCA.}
\label{fig4}
\end{figure*}
To support this prediction, we conducted experiments with PSCs fabricated by using MAPI, TCA and carbon electrode, where the concentration of TCA was changeable like 0, 5, 10, 15, 20 mg/mL.
As shown in Figure~\ref{fig4}a for XRD patterns, the perovskite film with 5 mg/mL TCA exhibited the peaks with weaker intensities including (220) and (310), indicating its weakened crystallization, whereas the enhanced crystal growth in the [110] direction was found for the perovskite with 15 mg/mL TCA.
SEM images for the perovskite films without TCA (Figure~\ref{fig4}e) and with 15 mg/mL TCA (Figure~\ref{fig4}f) indicate that some amount of precursor solution was permeated into the mesoporous scaffolds to form small-sized crystals, while the remaining precursor solution formed rod- and ball-shaped crystals for the former and latter cases on the surface of \ce{TiO2} films.
Note that the ball-shaped particles with a nanometer size were prone to penetrate into the scaffolds, resulting in the enhancement of crystallinity, while it is difficult for the rod-shaped crystals.
As shown in Figure~\ref{fig4}b, the light harvesting capability of the perovskite films was enhanced by adding and increasing the concentration of TCA, due to higher absorbance values for higher concentration especially at the longer wavelength region over 500 nm.
The PL spectra indicates that TCA was not incorporated into the MAPI lattice due to their identical peak positions of 760 nm for the TCA concentrations of 0 and 15 mg/mL, and the non-radiative charge recombination could be prohibited effectively by adding TCA due to higher PL intensity for 15 mg/mL concentration (Figure~\ref{fig4}c).
From the $J-V$ curves in Figure~\ref{fig4}d, the highest PCE of 13.03\% was found in the device with 15 mg/mL TCA, which is $\sim$3\% higher than the device without TCA.

In summary, this study illuminates fundamental aspects underlying the impact of perovskite interface with Lewis base and graphene on the performance of solar cell device.
It reveals stronger binding of $S$-donors thiocarbamide and thioacetamide with the MAPI(110) surface and graphene than $O$-donor dimethyl sulfoxide.
The interface dipole directing from the perovskite to graphene induces work function reduction, indicating the promotion of electron extraction by graphene, and they tend to increase going from TCA to DMSO.
The DFT-based formation energies of surface defects provide an evidence of deep trap passivation for the TCA-related perovskite interface in line with the experimental observation of performance enhancement in solar cells prepared using TCA with different concentrations.

\section*{Acknowledgments}
This work was supported partially by the State Committee of Science and Technology, Democratic People's Republic of Korea.
The calculations have been performed on the HP Blade System C7000 (HP BL460c) that is owned and managed by the Faculty of Materials Science, Kim Il Sung University.

\section*{Appendix A. Supplementary data}
Supplementary data related to this article can be found at URL.

\section*{\label{note}Notes}
The authors declare no competing financial interest.

\bibliographystyle{elsarticle-num-names}
\bibliography{Reference}

\end{document}